# SIZE EFFECTS IN RADIOSPECTROSCOPY SPECTRA OF FERROELECTRIC NANOPOWDERS

M. D. Glinchuk, I. V. Kondakova, A. M. Slipenyuk,
I. P. Bykov, V. V. Laguta, A. V. Ragulya, V. P. Klimenko

*Institute for Problems of Materials Science, NASc of Ukriane,
Krjijanovskogo, 3, 03680 Kiev-142, Ukraine*

The theoretical and experimental investigation of ferroelectric nanopowders is performed. The manifestation in radiospectroscopy spectra of size driven ferroelectric–paraelectric phase transition at some critical particle average size $\bar{R} = R_c$ was the main goal of the consideration. In theoretical part the size effect for the materials with ferroelectric tetragonal phase at room temperature and cubic paraelectric phase was considered allowing for the spontaneous polarization inhomogeneity inside a particle and distribution of particle sizes. In ESR the transformation of the spectra from tetragonal symmetry to cubic symmetry lines with decreasing of nanoparticle sizes was calculated. The method of $R_c$ value extraction from the ratio of the different symmetry lines intensities in the absorption spectra was proposed. Measurements of $Fe^{3+}$ ESR spectra in nanopowder of $BaTiO_3$ were carried out at room temperature. The samples were prepared by rate-controlled method with different particle sizes, which depend on annealing temperature. The decrease of intensity of tetragonal symmetry ESR lines of $Fe^{3+}$ and appearance of cubic symmetry line with asymmetry of the shoulders was observed with the average sizes decrease with complete disappearance of tetragonal spectrum at $\bar{R} \leq 40$ nm. The comparison of the theory with experiment was carried out. The theory fits experimental data pretty good. The value of critical size $R_c \approx 40$ nm was extracted from ESR data. The asymmetry and broadening of right hand side shoulder of ESR cubic symmetry line was shown to be related to contribution of paramagnetic centers in the vicinity of the particles surface with lower than cubic symmetry. The deconvolution of the cubic line allowed to show, that this region size is about 3 nm.
PACs: 73.22.-f, 61.72.Hh, 68.35.Rh

## 1. Introduction

The nanomaterials attract much attention of the scientists and engineers because of their unusual physical properties related to the size effects [1-3], which made these materials prospective for various technical applications (see [4] and ref. therein]. In ferroelectric nanomaterials the most prominent size effect was shown to be size driven ferroelectric-paraelectric phase transition, which takes place at some critical nanoparticle size or film thickness. The existence of a critical size of ferroelectric nanoparticles at arbitrary temperature was shown by various experimental methods, including X-ray diffraction, transmission electron microscopy, dielectric spectroscopy and radiospectroscopy [6-8] however with strong scattering of critical size values. The development of the method of critical size true value measurement stays one of the most import problems up to now.

Application of radiospectroscopy methods for nanomaterials investigation was shown to be especially useful in investigation of nanomaterials properties (see e.g. [9-11]). The main advantage of these methods is their ability to study local properties of the solids that is very important for nanomaterials, namely nanograin ceramics, nanoparticle powders, thin films. It is because the nanomaterials properties are known to be inhomogeneous changing from those on the surface to those in the bulk [12-14]. So that the investigation of local properties rather than averaged ones has to be important. On the other hand radiospectroscopy methods contrary to dielectric spectroscopy methods do not need any electrodes on the samples, so that nanopowders can be studied. Allowing for width of the nanoparticle sizes distribution in powders can be smaller than in nanograin ceramics the information about size effects extracted from measurements of the powders can be more precise than that obtained from nanograin ceramic investigation [15].

Therefore the radiospectroscopy can be considered as the sensitive method for the size effects investigation in nanomaterials. The basis of the method of these effects investigation has to be the theoretical description of radiospectroscopy spectra peculiarities originated from the size effects allowing for characteristic features of the material. In particular the disappearance of spontaneous polarization at critical size in ferroelectric nanomaterial that must lead to the change of symmetry, the inhomogeneity of the properties and the distribution of the nanoparticle sizes, which have to lead into inhomogeneous broadening of the resonance lines.

In this paper the results of experimental and theoretical investigation of $BaTiO_3$ nanopowder are presented. The transformation of ESR spectra of $Fe^{3+}$ paramagnetic centers symmetry from tetragonal to cubic with the asymmetry of right hand side shoulder under decrease of the particles sizes was observed and described theoretically allowing for polarization inhomogeneity and its dependence on the sizes distribution. The asymmetry of cubic lines was explained by influence of surface tension on the spectra via spin-phonon interaction. The value of critical radius extracted from experimental data appeared to be 40 nm.

## 2. Theory

In what follows we will study the crystalline field size effect originated from spontaneous polarization $P$ in ferroelectric phase and surface tension influence on cubic symmetry line in paraelectric phase of perovskite structure. Keeping in mind that axial symmetry crystalline field constant $D$ for ESR spectra is known to be proportional to $P$ or to $P^2$ for paramagnetic centers without or with inversion center respectively and quadrupole coupling constant related to crystalline field gradient is proportional to $P^2$ let us begin with consideration of inhomogeneous polarization in nanoparticles.

### 2.1. Spontaneous polarization in ferroelectric nanoparticles

The inhomogeneity of spontaneous polarization originated from its value change from that on the surface to that in the bulk lead to necessity of consideration in phenomenological theory approach the free energy functional with gradient of polarization and surface energy contribution (see e.g. [16]). Equilibrium value of polarization can be found by variation of the functional that lead to Euler-Lagrange equation for the polarization with boundary condition originated from surface energy. For the spherical particles with size $R$ one has to solve the equation [12]

$$aP + bP^3 + cP^5 - \delta\left(\frac{d^2P}{dr^2} + \frac{2}{r}\frac{dP}{dr}\right) = a_0 \bar{P}; \tag{1a}$$

$$\left(\lambda \frac{dP}{dr} + P\right)_{r=R} = 0. \tag{1b}$$

Here $\bar{P} = 1/V \int P(r) dV$ ($V$ is particle volume), the coefficient $a$ depends on temperature, surface tension coefficients, particle size and other material characteristics (see [12] for details), while the other coefficients in Eq. (1a) are independent on $T$, $R$ and $r$ is the length of radius-vector that describes the position of the point inside the particle. In the boundary condition (1b) $\lambda$ is extrapolation length. The application of direct variational method for solution of Euler-Lagrange equation for nanoparticles similarly to what was proposed for thin ferroelectric films earlier [17] allows to write polarization as [12]:

$$P(r, R) = P_V \left(1 - \frac{R}{r}\frac{sh(r\sqrt{a/\delta})}{M(R)}\right);$$

$$M(R) = \lambda\sqrt{\frac{a}{\delta}} ch\left(R\sqrt{\frac{a}{\delta}}\right) + \left(1 - \frac{\lambda}{R}\right) sh\left(R\sqrt{\frac{a}{\delta}}\right). \tag{2}$$

Here the variational parameter $P_V$ can be obtained by minimization of the free energy of conventional form, but with renormalized coefficients:

$$F = A_R \frac{P_V^2}{2} + B_R \frac{P_V^4}{4} + C_R \frac{P_V^6}{6}. \tag{3}$$

The renormalized coefficients can be written approximately in the following form:

$$A_R \approx \alpha(T - T_{cr}(R)) \tag{4a}$$

or

$$A_R \approx \alpha(T - T_c)\left(1 - \frac{R_c(T)}{R}\right) \tag{4b}$$

$$R_c \approx \frac{R_c(T=0)}{1 - T/T_c}$$

$$B_R \approx b, \quad C_R \approx c, \tag{4c}$$

where $T_{cr}(R)$ and $R_c(T)$ are respectively the temperature and radius of size driven ferroelectric–paraelectric phase transition with ferroelectric phase at $T \leq T_{cr} \leq T_c$, $R \geq R_c(T)$, $T_c$ and $\alpha$ are respectively paraelectric–ferroelectric phase transition temperature and inverse Curie-Weiss constant in the bulk ferroelectric.

It follows from Eq. (3), that parameter $P_V(R)$ for the phase transition of the second or the first order can be written in conventional for bulk ferroelectric form respectively as

$$P_V^2 = -\frac{A(R)}{B(R)} \quad \text{or} \quad P_V^2 = -\frac{B_R}{2C_R}\left(1 + \sqrt{1 - \frac{4 A_R C_R}{B_R}}\right) \tag{5}$$

The Eqs. (2), (4), (5) allow to write the dependence of the polarization on $r$ and $R$ for the phase transition of the first and second order.

### 2.2. Spectra of magnetic resonances

The shape of homogeneously broadened ESR and NMR lines depends on the main mechanism of the line broadening and in general case it can be represented in Gaussian, Holtczmarkian or Lorentzian form [18]. Let us suppose that it can be described in the form of Gaussian, so that

$$I(\omega, R, r, \theta) = \frac{1}{\sqrt{2\pi}\Delta} \exp\left(-\frac{(\omega - \omega_0(r, R, \theta))}{2\Delta^2}\right), \tag{6}$$

where $\omega_0(r,R,\theta)$ is resonance frequency, $\theta$ is the angle that reflects orientation of paramagnetic center symmetry axes relatively constant magnetic field, $\sqrt{2\ln 2}\Delta$ is half-width on half-height, the distribution of $R$, $r$ and $\theta$ will lead to inhomogeneous broadening of the lines.

In what follows we will consider ferroelectric materials with perovskite structure, that has cubic and tetragonal symmetry in paraelectric and ferroelectric phase respectively, the symmetry lowering being related to polarization. So that $\theta$ is the angle between external magnetic field and polarization (for tetragonal symmetry crystalline field) or electric field gradient principal axes. However even in paraelectric phase for paramagnetic centers in the vicinity of surface the symmetry has to be lower than cubic because the absence of inversion center near the surface and so only axis or planes normal to the surface conserve out of all symmetry elements of the crystal, what leads to essential changes in the form of spin-Hamiltonian [19]. Let us suppose that resonance frequency in ESR case is defined by crystalline field. In ferroelectric phase axial constant $D$ is originated from polarization and $D \sim P^2(r,R)$ or $D \sim P(r,R)$ respectively for paramagnetic center with or without inversion center. The size effect for crystalline field constant in paraelectric phase was considered recently in core and shell model [10], where shell is the region of $\Delta R$ size in the vicinity of the particle surface, where paramagnetic centers "feel" the influence of surface, and core (of $R - \Delta R$ size) has the symmetry of the bulk, i.e. cubic symmetry in the considered ferroelectrics with perovskite structure. In the case of NMR we will consider nuclei with large quadrupole moment that has to interact with electric field gradient $V$. In ferroelectric phase $V_{zzf} \sim P^2(r,R)$ and in paraelectric phase $V_{zzp}$ is originated from the shell with lower than cubic symmetry. Qualitatively it is obvious that contribution of shell both in ESR and NMR has to be larger for smaller particles because of increase of surface to volume ratio. Quantitatively we calculated it earlier [10] by taking into account the influence of hydrostatic pressure $p = 2k/R$ ($k$ is coefficient of surface tension) on resonance frequency via spin-phonon coupling constant. Because resonance field depends on orientation of the sample in the external magnetic field via matrix elements of paramagnetic probe spin operators (e.g. axial crystalline field constant $D = \hat{S}_z^2 - 1/3 S(S+1)$, where $\hat{S}_z$ is spin operator, $S$ is spin value) the powder spectra we are interested in can be obtained by averaging over angle θ. In the case of nanoparticle powders the sizes of nanoparticles are obligatory distributed so that the intensity of magnetic resonance spectra can be written as

$$I(\omega) = \frac{3}{2\sqrt{2\pi}\Delta} \int_0^\pi \sin\theta\, d\theta \int_0^\infty \frac{dR}{R^3} f(R) \int_0^R r^2 dr \exp\left(-\frac{(\omega - \omega_0(r, R, \theta))^2}{2\Delta^2}\right) \qquad (7)$$

Here $f(R)$ is distribution function of the particle size. In what follows we will supposed the Gaussian form of this distribution function, that has the form [15]

$$f(R) = \frac{2}{\sigma\sqrt{\pi}\,(erf(R_0/\sigma) + 1)} \exp\left(-\frac{(R - R_0)^2}{\sigma^2}\right) \qquad (8a)$$

$$\bar{R} = R_0 + \frac{\sigma \exp(-R_0/\sigma)^2}{\sqrt{\pi}\,(1 + erf(R_0/\sigma))} \qquad (8b)$$

where $R_0$ and $\sqrt{\ln 2}\,\sigma$ are the most probable value of radius and half-width at half-height.

Note, that in real nanopowder samples distribution function of sizes can have more complex form than Gaussian (8), even with two maxima, as one can see later. For the sake of illustration here we took the simplest form of $f(R)$ given by Eq. (8) although it is more correct to take experimentally observed form for $f(R)$ as we shall do later for comparison the theory with experiment.

It is convenient to begin with averaging over $d\theta$ to obtain powder spectra. It is known that the most intensive lines in this spectra belong to $+1/2 \leftrightarrow -1/2$ transition independently on spin value [20], the line shape and positions being calculated numerically in the majority of the cases (see e.g. [21]). The analysis of the calculated powder spectra has shown that the central transition spectra for axial symmetry can be approximately represented in the form of two Gaussians, which correspond to effective $g$-factors, namely to $g_\parallel$ and $g_\perp$, and by one Gaussian for cubic symmetry. Because of this in what follows in Eq. (7) we will drop the integration over $d\theta$ and substitute two exponents for the single exponent in ferroelectric phase, that reflects the axial symmetry and conserve one exponent in paraelectric phase.

Keeping in mind, that because of aforementioned reasons resonance frequency has to be different in ferroelectric ($R_c \leq R < \infty$) and paraelectric phase ($0 \leq R \leq R_c$) the Eq. (7) can be presented as

$$I(\omega) = \frac{1}{\sqrt{2\pi}\Delta} \frac{3}{2} \left\{ \int_{R_c}^\infty \frac{dR}{R^3} f(R) \int_0^R r^2 dr \left[ \frac{2}{3} \exp\left(-\frac{\omega - q_\perp \omega_{0f}(r, R)}{2\Delta^2}\right) \right. \right.$$
$$\left. + \frac{1}{3} \exp\left(-\frac{\omega - q_\parallel \omega_{0f}(r, R)}{2\Delta^2}\right) \right] + \int_0^{R_c} dR\, f(R) \left[ A_c(R) \exp\left(-\frac{(\omega - \omega_{0pc})^2}{2\Delta_{1c}^2}\right) \right.$$
$$\left. \left. + A_s(R) \exp\left(-\frac{(\omega - \omega_{0ps})^2}{2\Delta_{1s}^2}\right) \right] \right\} \qquad (9)$$

where $A_c(R) = \dfrac{(1 - \Delta R/R)^3}{\sqrt{2\pi}\Delta_c}$, $A_s(R) = \dfrac{1 - (1 - \Delta R/R)^3}{\sqrt{2\pi}\Delta_s}$ are the coefficients obtained in [10] from the condition of normalization, $\omega_{0pc}$ and $\omega_{0ps}$ are resonance frequency respectively for core and shell regions. Here resonance frequency in the ferroelectric phase $\omega_{0f}$ can be proportional to linear or squared polarization in ESR case and to the fourth power of polarization in NMR case because for the transition +1/2 ↔ –1/2 the resonance frequency is proportional to square quadrupole coupling constant, namely $\omega_{0q} \sim (eQ)^2 P^4(r,R)$, where $eQ$ denotes the quadrupole moment of nuclei.

In the second integral in Eq. (9), that corresponds to paraelectric phase, resonance frequency in shell and core regions can be written respectively as $\omega_{0ps} = \omega_{0c} + \eta/R$ ($\eta$ is proportional to surface tension and spin-phonon coefficient) and $\omega_{0pc} = \omega_{0c}$ with $\omega_{0c}$ value close to resonance frequency in the bulk paraelectric phase of cubic symmetry. In general case one can expect the appearance of two separate lines, namely for core and for shell region. This depends on the difference $\Delta\omega = \omega_{0ps} - \omega_{0pc}$ value and the width of the line, in particular for $\Delta\omega / \sqrt{\Delta_{1c}^2 + \Delta_{1s}^2}$ less than 2 or 3 one can expect the line broadening rather than splitting. More precisely, allowing for g-factor of paramagnetic center on the surface is shifted to smaller value [19], the asymmetry and broadening of ESR line right hand side shoulder is expected. In NMR case the application of magic angle spinning that decrease the line width allows to observe the transformation of $^{17}O$ spectra in MgO defined by chemical shift mechanism from one narrow line into two broadened lines when decreasing the crystallite sizes from 35 nm to 2.5 nm [11]. All these features were successfully described in core-shell model earlier [10].

In the case of NMR paramagnetic probe with large quandupole moment only shell region can contribute to electric field gradient so that the line broadening can be observed instead of narrowing, expected for cubic symmetry of paraelectric phase. So that it is not clear apriory if essential difference of quadrupolar nuclei NMR line in ferroelectric nanopowder with transition from tetragonal ferroelectric to cubic paraelectric phase with the particles sizes decreasing can be expected.

*2.3. Size effect in calculated resonance spectra*

In this section we will represented the form of resonance spectra calculated on the basis of general formula of previous section for the parameters close to $BaTiO_3$ ferroelectric material.

The ESR spectra for several nanopowder samples with the ratio of the most probable value of particle size to critical size $R_0/R_c = 5; 3; 2.1; 1.2$ (the ratio of $\overline{R}/R_0$ can be obtained on the basis of Eq. (8b)) are represented as the curves 1, 2, 3, 4 respectively in all the Figs. in supposition that contribution of shell region to cubic line can be neglected. We will discuss this contribution later in the chapter devoted to the comparison with experiment.

In Fig. 1 ESR spectra of paramagnetic probe with inversion center ($D \sim P^2$) for ferroelectric nanopowders with the I$^{st}$ order phase transition are depicted for the cases of $\sigma = \frac{1}{2}R_c$ (a) and $\sigma = R_c$ (b). One can see that with the decrease of the particle sizes two lines of axial symmetry spectra became less intensive and broader so that there is only one broad line substituted for two lines (see curve 4) and they are shifted to lower magnetic field, where cubic symmetry line arised, its intensity increases with the sizes decrease. The comparison of Figs. 1a and 1b shows, that broader distribution function of sizes leads to the appearance of cubic line for larger $R_0/R_c$ value (compare curves 2) because the probability to find the particles with $R_0 \leq R_c$ increases as well as $\bar{R} - R_0$ value (see Eq. (8b)) with σ increasing. However the intensity of cubic line increases more essentially with the sizes decrease. For the sake of comparison in Fig. 2 we gave the view of the spectra for the phase transition of the II$^{nd}$ order. It follows from the comparison of Figs. 2a and 2b that although qualitatively they are similar, quantitatively size effect (axial lines shift and broadening) is more pronounced for the paramagnetic centers with inversion center. One can see from comparison of the corresponding curves in Figs. 2a and 1b, that size effects are stronger for the ferroelectric nanopowders with II$^{nd}$ order than for the I$^{st}$ order phase transitions.

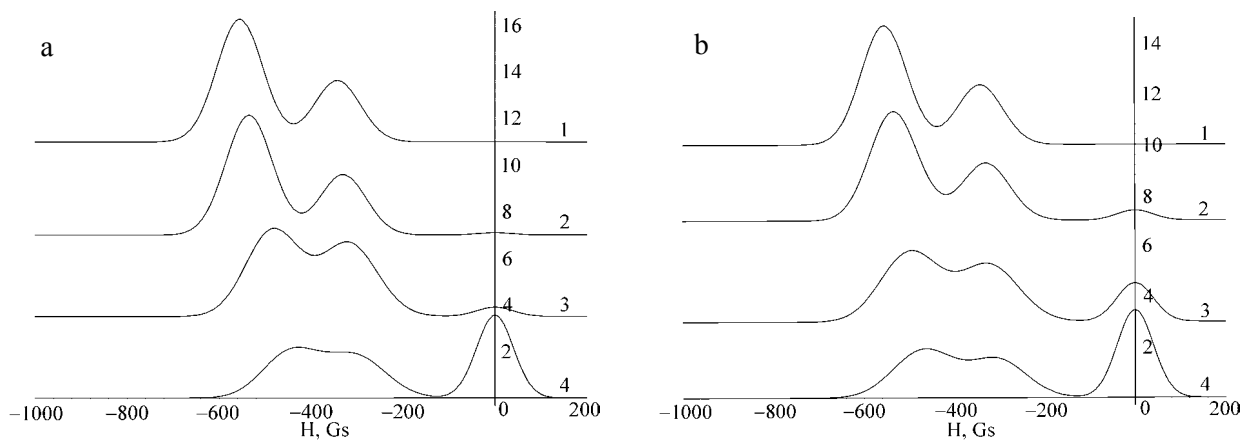

**Figure 1.** Calculated ESR absorption spectra of paramagnetic probe with inversion center for several nanopowder ferroelectrics with I$^{st}$ order phase transition. The parameters of size distribution function are the following: $R_0/R_c = 5$ (1), 3 (2), 2.1 (3), 1.2 (4); $\sigma/R_c = 0.7$ (a), 1.4 (b).

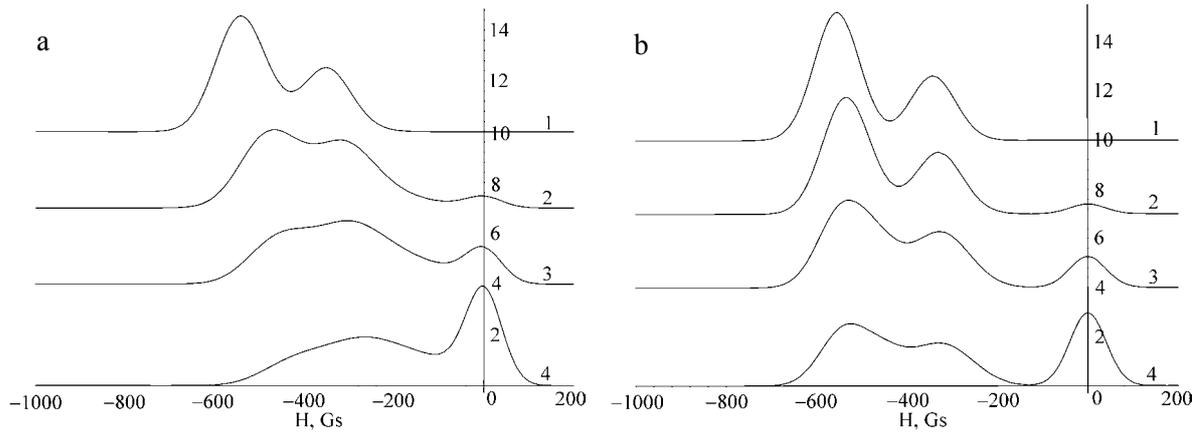

**Figure 2.** Calculated ESR absorption spectra of paramagnetic probe with (a) and without (b) inversion center for several nanopowder ferroelectrics with II$^{st}$ order phase transition. The parameters of distribution function are the same as in Fig. 1,b.

To find out how the form of distribution function influence resonance spectra we depicted them in Fig. 3 for the same parameters as those in Fig. 1b, but for the curves 3 and 4 $f(R)$ was taken in the form of two Gaussians respectively with $R_{01}/R_c$ = 1.4 and 1.2; $R_{02}/R_c$ = 4.5 and 3.8 with $\sigma_1 = \sigma_2 = 0.3R_c$. The comparison of Figs. 1b and 3 shows that two Gaussian distribution function leads to broadening of all the lines and stronger smearing of two lines axial spectra.

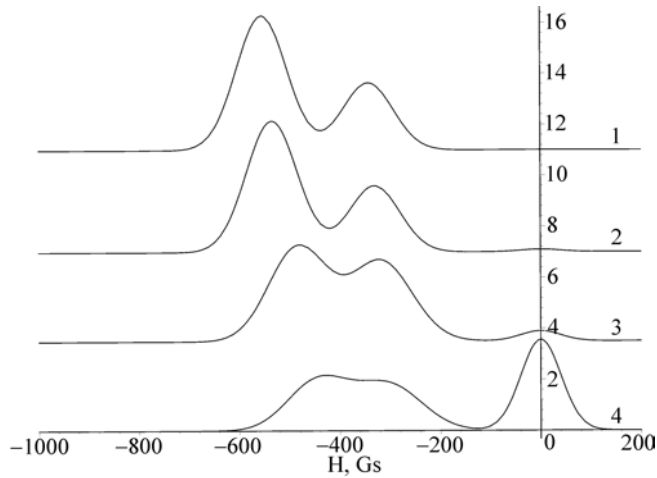

**Figure 3.** Calculated ESR absorption spectra of paramagnetic probe with inversion center for nanopowder ferroelectrics with I$^{st}$ order phase transition and size distribution function in the form of one Gaussian (for curves 1 and 2) and of two Gaussians (for curves 3 and 4).

The calculations had shown that inhomogeneity of polarization, which increases with the sizes decrease (see Eq. (2)), contributes essentially into smearing and disappearance of axial lines for small enough particles. On the other hand for the sizes $R/R_c \geq 3$, which correspond to the curves 1 and 2 in the Figs. polarization is nearly homogeneous with accuracy of several percent.

One can see that in all the considered cases the decreasing and broadening of axial symmetry spectra and increasing of cubic line was the main feature of size effect. Qualitatively this can be expected because of the increase of number of the particles with $R < R_c$ when the size of particles decreases, the rate of cubic line increase being dependent on the value of $R_c$. Quantitatively the value of $R_c$ can be extracted from observed spectra namely from ratio $I_t/I_c$ of the integral intensity of axial and cubic absorption lines as the function of $R_0/R_c$ (see Fig. 4). It is seen that to obtain $R_c$ value one has to know $I_t/I_c$ and the parameters $R_0$ and $\sigma$ of particle sizes distribution function, which are the main characteristics of a nanopowder sample. It is worth to underline that the proposed method of $R_c$ value extraction the absorption spectra has to be recorded, because the accuracy of integral intensity calculation from the derivative of powder samples spectra can be very low.

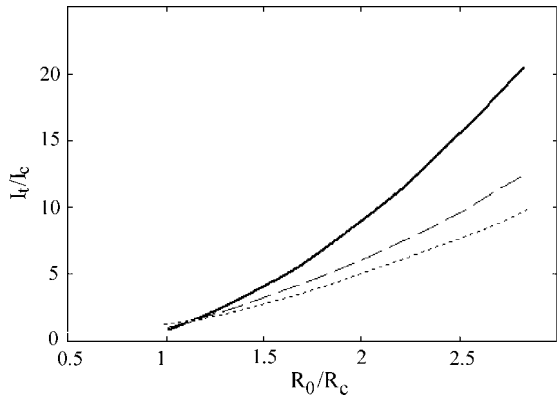

**Figure 4.** Size dependence of ratio of the integral intensities of axial and cubic absorption lines for different parameters of distribution function $R_0/\sigma$: 2.4 (dotted line); 2.8 (dashed line); 3.5 (solid line).

### 3. Experiment
#### 3.1. Sample preparation

The nanocrystalline barium titanate powder doped with iron (0.01 mol%) had been prepared through the oxalate technology. The barium titanyl-oxalate with iron additive $BaTi_{0.999}Fe_{0.001}O(C_2O_4)_2 \cdot nH_2O$ was used as a precursor and had been synthesized by wet-technology from the following raw-materials: titanium tetrachloride, barium nitrate, oxalic acid and iron chloride. All chemicals were of high chemical purity. The solutions of $TiCl_4$, $FeCl_3$ and $Ba(NO_3)_2$ were added drop vise to the oxalic acid water solution at thorough stirring. Precipitation of the doped oxalate was carried out at pH = 1 and 15°C. Obtained suspend was washed with distilled water, filtered and dried at 100°C in air. Obtained barium titanyl-oxalate was annealed at the temperatures of 900–1200°C with heating rate of 50°C/h to produce powders with different size of the particles.

The size of the particles was detected using ZETASIZER 1000 HS/3000 HS equipment. It was shown that the average size decrease from several hundreds nm to several tens nm with the decreasing of annealing temperature from 1350°C to 900°C. Analysis had shown that one or two peaks in the size distribution were observed for all powders. For the sample annealed at 900°C the position of two peaks was $\bar{R} \approx 40$ nm and $\bar{R} \approx 140$ nm.

The XRD apparatus (Dron 3M, USSR) was used to analyze phase composition of the obtained samples. It was shown that samples, which have been annealed below 1000°C, were predominantly of cubic symmetry (with tracks of tetragonal symmetry of very low intensity) barium titanate. Samples, which have been annealed above 1000°C, were of tetragonal symmetry.

Electron Spin Resonance (ESR) spectra were recorded at room temperature on a spectrometer operating at microwave frequency 9 GHz.

#### 3.2. Experimental rezults

In Fig. 5 ESR spectra observed in $BaTiO_3$ powder samples with different size of nanoparticles recorded at room temperature have been presented. Because all samples were doped by iron we supposed that the observed lines mainly originate from $Fe^{3+}$ ions. The belonging of the measured resonances to $Fe^{3+}$ impurity was confirm on the base of the simulation of powder spectrum using spin Hamiltonian parameters found from single-crystal measurements and reported, for instance, in Ref. [22]. In the simulation procedure we used numerical diagonalization of the spin-Hamiltonians that allows us to determine the resonance magnetic fields of all transitions (including forbidden) as well as their probabilities. In Fig. 5 the simulated powder spectrum of $Fe^{3+}$ ion in the tetragonal phase of $BaTiO_3$ (see *f*) is shown. One can see, that the spectrum *e* for the largest particles (about of micron size) contains all the transitions of a bulk sample so that spectrum *e* will be considered as that for a bulk with tetragonal symmetry. It follows from the other samples spectra that, cubic symmetry line at $B \sim 3350$ Gs appears in $Fe^{3+}$ powder spectrum and its intensity increases with the particles size decrease (see *d*, *c*, *b*, *a* spectra). The comparison of these spectra shows also the decrease and broadening of the most intensive lines in powder tetragonal symmetry spectra, which correspond to central $+1/2 \leftrightarrow -1/2$ transition, while the lines of the other fine structure transitions are practically absent in spectra *c*, *b*, *a*.

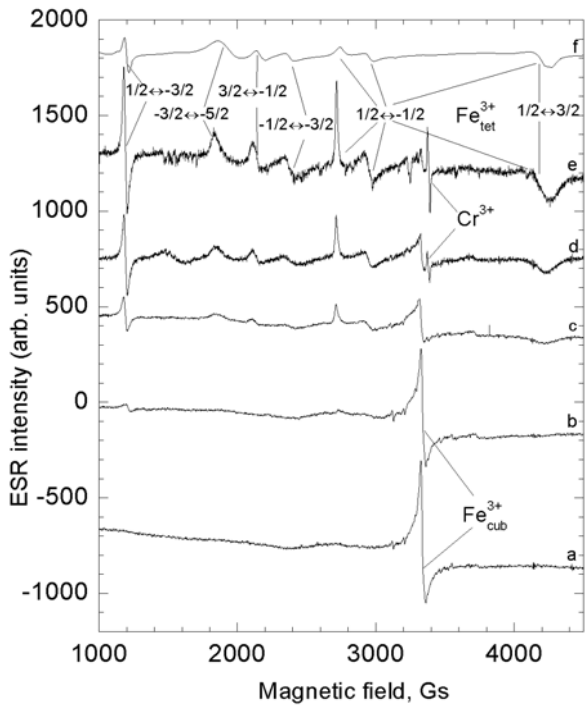

**Figure 5.** ESR spectra at $T = 20°C$ in nanopowder $BaTiO_3$: 0.01 % Fe samples annealed at different temperatures: a – 900°C, b – 1000°C, c – 1100°C, d – 1200°C, e – 1350°C; f – calculated spectrum in tetragonal phase of $BaTiO_3$.

It is seen, that finally there is only cubic symmetry line in spectrum for sample *a*, that was annealed at the lowest temperature $T = 900°C$. This is in agreement with X-ray data about predominately cubic symmetry of barium titanate in this sample. The peculiar feature of the cubic line (see spectrum *b* and *a*) is its asymmetry with different width of the left and right hand side shoulders. One of the reasons of this asymmetry can be peculiarities of cubic line powder spectra (see Fig. 6). It is seen from this Figure that beside narrow intensive line the small intensity lines have to be in its vicinity, these lines being more intensive on left hand side. It obvious that these lines are residuals of tetragonal symmetry fine transitions in powders because they are absents in single crystals. Since the lines of tetragonal symmetry fine transitions are practically absent on right hand side of cubic line (see spectra *e*, *d*) and as a result the intensity of right hand side shoulder in spectrum (see Fig. 6) is very small it has to be another reason for the broadening of right hand side shoulder in the observed spectra. We will discuss it later in the next section.

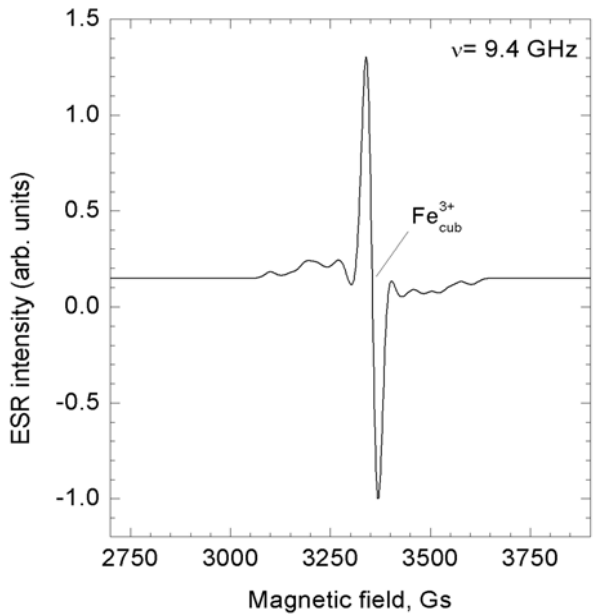

**Figure 6.** Calculated ESR spectrum for $Fe^{3+}$ ions in cubic phase of $BaTiO_3$ powders.

### 4. Comparison of the theory with experiment

The theoretical calculations performed in Sec. 2 have shown, that with decreasing of the particles size the intensity of tetragonal symmetry central transition decreases, its width increases and cubic symmetry lines arises, its intensity increases with the sizes decrease. This behaviour in good agreement with the transformation the observed spectra described earlier. Because in Sec. 2 we calculated absorption spectra here in Fig. 7 we represented the derivative of powder spectra like what was observed experimentally.

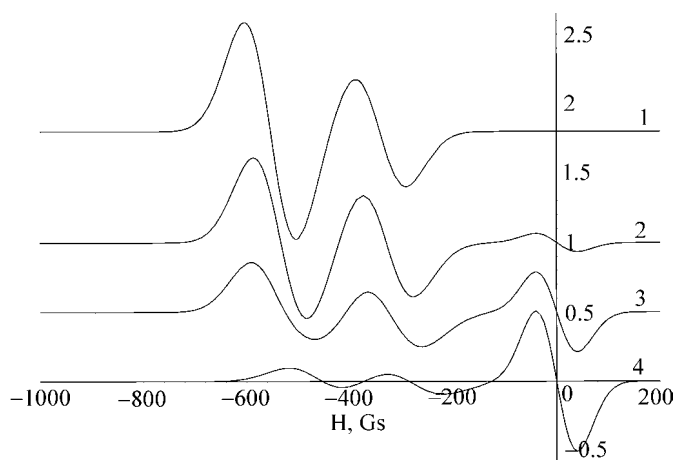

**Figure 7.** Calculated ESR spectra in the form of derivatives for powder samples with the size distribution function in the form of one Gaussian (for curves 1 and 2) and of two Gaussians (for curves 3 and 4) with parameters $R_0/R_c = 5$ (1), 3 (2); $R_{01}/R_c = 2.1$ and $R_{02}/R_c = 4.5$ (3); $R_{01}/R_c = 0.9$ and $R_{02}/R_c = 2.8$ (4).

In accordance with experiment we took for spectrum 4 the sizes distribution function in the form of two Gaussians with the parameters $R_{01}/R_c = 0.9$ and $R_{02}/R_c = 2.8$ allowing for small enough $R_0$ value $\bar{R} > R_0$ [15]. We decreased these parameters in comparison with those in Fig. 3 to decrease the intensity of tetragonal lines. One can see, that in spectrum 4 the intensity of tetragonal lines are practically negligibly small similarly to observed spectra a, b. Allowing for that $\bar{R}_1 = 40$ nm for spectrum a one can conclude that $R_c \approx 40$ nm although the value $\bar{R}_2 \leq 140$ nm lead to a little bit larger value. Note, that because both in theory and in experiment there is some residuals of tetragonal symmetry lines the real critical size may be smaller than 40 nm. Let us proceed to consideration of the nature of asymmetry and broadening of right hand side shoulder of cubic line. It what follows we will show that its origin is the contribution of the paramagnetic centers in the vicinity of the particle surface, i.e. in the particle shell.

As it was shown in [19] the difference in the g-factors values corresponding to paramagnetic centers from the core and shell is $\Delta g = 5 \cdot 10^{-3}$ i.e. ESR line originated from the centers on the particles surface has to be shifted to the larger magnetic field region relatively bulk line. Because of the asymmetry of the form of observed ESR line we assume that it consists from several lines. The deconvolution of the spectra was made in Peak Fit program with the assumption that the contributions from core and shell are comparable. The attempt to deconvolute spectra into too line was not successful – it was necessary to add the third line. We suppose that the source of this additional contribution to the ESR spectrum can be the residuals from tetragonal symmetry lines as we discussed in previous section (see also Fig. 6). In Fig. 8 ESR spectra of the samples which have been annealed at 900°C and 1100°C as well as the results of the spectrum deconvolution and total spectrum are presented. It is seen that in both spectrum narrow (~1.9 mT) line with g-factor value 2.003 presents. This line corresponds to cubic $Fe^{3+}$ spectrum in the bulk samples (single crystals and ceramics). The line of small intensity with g-factor value shifted to the larger magnetic field side originates from the centers near the surface of the nanoparticles. The intensity of this line increases with the temperature of the samples annealing decrease that corresponds to the increasing of the surface influence on the powders properties. The third broad line in the spectrum deconvolution comprises the contributions from the tetragonal lines residuals. Theoretical description of the ESR lines corresponding to core (line 1) and shell (line 2) gave us possibility to determine the size of shell i.e. the size of the region that is sensitive to the surface influence. The obtained value of $\Delta R$ is approximately 3 nm. The details of the calculations which lead to this value will be published elsewhere.

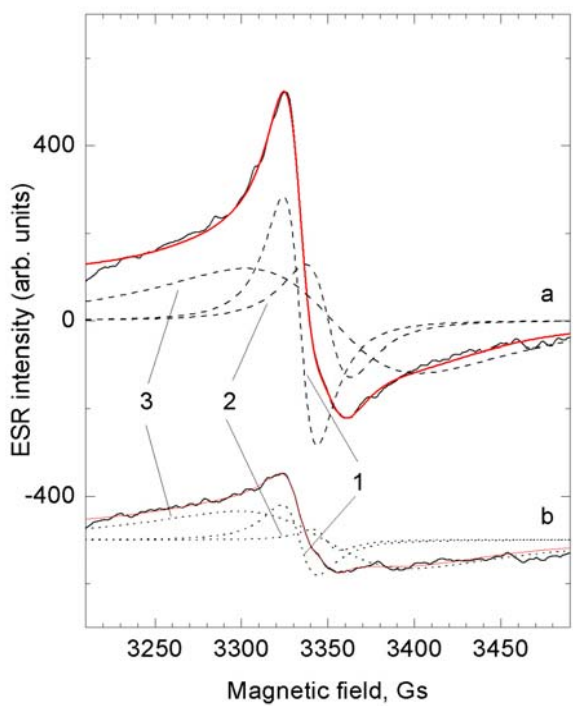

**Figure 8.** ESR spectra (solid lines) in nanopowders $BaTiO_3$ annealed at 900°C (a), 1100°C (b). Dash lines – the results of the spectrum deconvolution correspondingly: 1 – core, 2 – shell, 3 – the residuals of tetragonal lines.